\documentclass[12pt]{cernart}

\usepackage{epsfig}
\usepackage{graphicx}
\usepackage{amsmath,amssymb}

\include{defin}
\usepackage{amssymb}
\newcommand{\be} {\begin{equation}}
\newcommand{\ee} {\end{equation}}
\newcommand{\bea} {\begin{eqnarray*}}
\newcommand{\eea} {\end{eqnarray*}}
\newcommand{\kppg}{K^{\pm} \to \pi^{\pm}\pi^{0}\gamma}
\newcommand{\knppg}{K_{L,S} \to \pi^{+}\pi^{-}\gamma}

\newcommand{\pio}{\pi^{0}}
\newcommand{\kpppn}{K^{\pm}\to\pi^{\pm}\pi^{0}\pi^{0}}
\newcommand{\kppp}{K^{\pm}\to\pi^{\pm}\pi\pi}

\newcommand{\kpp}{K^{\pm}\to \pi^{\pm}\pi^{0}}
\newcommand{\ecmk}{T^{*}_{\pi}}
\newcommand{\kspiopio}{K^0_S \to  \pi^{0} \pi^{0}}

\begin{document}

\begin{titlepage}
\begin{flushright}
CERN-PH-EP/2010-006 8 February 2010
\end{flushright}
\title{\bf \Large Measurement of the direct emission and interference terms and search for CP violation in the decay $\kppg$}
\begin{center}
{\Large The NA48/2 Collaboration}\\
\vspace{2mm}
 J.R.~Batley,
 G.~Kalmus,
 C.~Lazzeroni$\,$\footnotemark[1],
 D.J.~Munday,
 M.W.~Slater$\,$\footnotemark[1],
 S.A.~Wotton \\
{\em \small Cavendish Laboratory, University of Cambridge,
Cambridge, CB3 0HE, UK$\,$\footnotemark[2]} \\[0.2cm]
 R.~Arcidiacono$\,$\footnotemark[3],
 G.~Bocquet,
 N.~Cabibbo$\,$\footnotemark[4],
 A.~Ceccucci,
 D.~Cundy$\,$\footnotemark[5],
 V.~Falaleev,
 M.~Fidecaro,
 L.~Gatignon,
 A.~Gonidec,
 W.~Kubischta,
 A.~Norton$\,$\footnotemark[6],
 A.~Maier,\\
 M.~Patel,
 A.~Peters\\
{\em \small CERN, CH-1211 Gen\`eve 23, Switzerland}
% ...
\\[0.5cm]
%
\end{center}
\renewcommand{\thefootnote}{\fnsymbol{footnote}}
\footnotetext[1]{Corresponding authors, email: silvia.goy.lopez@cern.ch, mauro.raggi@lnf.infn.it}
\renewcommand{\thefootnote}{\arabic{footnote}}
\setcounter{footnote}{0}
\footnotetext[1]{University of Birmingham, Edgbaston, Birmingham,
B15 2TT, UK}
\footnotetext[2]{Funded by the UK Particle Physics and Astronomy
Research Council}
\footnotetext[3]{Dipartimento di Fisica Sperimentale
dell'Universit\`a e Sezione dell'INFN di Torino, I-10125 Torino,
Italy}
\footnotetext[4]{Universit\`a di Roma ``La Sapienza'' e Sezione
dell'INFN di Roma, I-00185 Roma, Italy}
\footnotetext[5]{Istituto di Cosmogeofisica del CNR di Torino,
I-10133 Torino, Italy}
\footnotetext[6]{Dipartimento di Fisica dell'Universit\`a e Sezione
dell'INFN di Ferrara, I-44100 Ferrara, Italy}
\footnotetext[7]{Scuola Normale Superiore, I-56100 Pisa, Italy}
\footnotetext[8]{CERN, CH-1211 Gen\`eve 23, Switzerland}
\footnotetext[9]{Faculty of Physics, University of Sofia ``St. Kl.
Ohridski'', 5 J. Bourchier Blvd., 1164 Sofia, Bulgaria}
\footnotetext[10]{Sezione dell'INFN di Perugia, I-06100 Perugia,
Italy}
\footnotetext[11]{Northwestern University, 2145 Sheridan Road,
Evanston, IL 60208, USA}
\footnotetext[12]{Centre de Physique des Particules de Marseille,
IN2P3-CNRS, Universit\'e de la M\'editerran\'ee, Marseille, France}
\footnotetext[13]{Department of Physics and Astronomy, George Mason
University, Fairfax, VA 22030, USA}
\footnotetext[14]{Dipartimento di Fisica, Universit\`a di Modena e
Reggio Emilia, I-41100 Modena, Italy}
\footnotetext[15]{Istituto di Fisica, Universit\`a di Urbino,
I-61029 Urbino, Italy}
\footnotetext[16]{Physikalisches Institut, Universit\"at Bonn,
D-53115 Bonn, Germany}
\footnotetext[17]{Funded by the German Federal Minister for
Education and research under contract 05HK1UM1/1}
\footnotetext[18]{SLAC, Stanford University, Menlo Park, CA 94025,
USA}
\footnotetext[19]{Royal Holloway, University of London, Egham Hill,
Egham, TW20 0EX, UK}
\footnotetext[20]{UCLA, Los Angeles, CA 90024, USA}
\footnotetext[21]{Laboratori Nazionali di Frascati, via E. Fermi,
40, I-00044 Frascati (Rome), Italy}
\footnotetext[22]{Institut de F\'isica d'Altes Energies, UAB,
E-08193 Bellaterra (Barcelona), Spain}
\footnotetext[23]{Funded by the German Federal Minister for Research
and Technology (BMBF) under contract 056SI74}
\footnotetext[24]{University College Dublin, School of Physics
Belfield, Dublin 4, Ireland}
\footnotetext[25]{Centro de Investigaciones Energeticas
Medioambientales y Tecnologicas, E-28040 Madrid, Spain}
\footnotetext[26]{Funded by the Austrian Ministry for Traffic and
Research under the contract GZ 616.360/2-IV GZ 616.363/2-VIII, and
by the Fonds f\"ur Wissenschaft und Forschung FWF Nr.~P08929-PHY}

\end{titlepage}

\begin{abstract}
We report on the measurement of the direct emission (DE) and interference (INT) terms of the $\kppg$ decay by the NA48/2 experiment at the CERN SPS.
From the data collected during 2003 and 2004 about 600k such decay candidates have been selected.
The relative amounts of DE and INT with respect to the internal bremsstrahlung (IB) contribution have been measured in the range $0<\ecmk<80$~MeV:
\begin{center}
$\mathrm{Frac_{DE}}{(0<\ecmk<80~\mathrm{MeV})}=(3.32\pm0.15_{stat}\pm0.14_{sys}) \times 10^{-2} $, \\
$\mathrm{Frac_{INT}}{(0<\ecmk<80~\mathrm{MeV})}= (- 2.35\pm0.35_{stat}\pm0.39_{sys}) \times 10^{-2} $ \\
\end{center}
where $\ecmk$ is the kinetic energy of the charged pion in the kaon rest frame.
This is the first observation of an interference term in $\kppg$ decays.

In addition, a limit on the CP violating asymmetry in the $K^+$ and $K^-$ branching ratios for this channel has been determined to be less than 1.5 $\times 10^{-3}$ at 90$\%$ confidence level.
\end{abstract}

\section{Introduction} \label{par:intro}
The decay channel $\kppg$ is one of the most interesting channels for studying the low energy structure of QCD. Radiative nonleptonic kaon decays, such as $\knppg$ and $\kpp$ are dominated by long distance contributions, but a small short distance contribution is also present and can be studied.

The total amplitude of the $\kppg$ decay is the sum of two terms: the inner bremsstrahlung (IB) associated with the $\kpp$ decay with a photon emitted from the outgoing charged pion, and the direct emission (DE) in which the photon is emitted at the weak vertex.
Using the Low theorem~\cite{Low} the branching ratio of the IB component can be predicted from that of the $\kpp$ channel, using QED corrections~\cite{christ, brib}.
As the $\kpp$ decay is suppressed by the $\Delta I=1/2$ rule, the IB component of $\kppg$ will be similarly suppressed, resulting in a relative enhancement of the DE contribution.

The DE term has been extensively studied in the framework of Chiral Perturbation Theory (ChPT)~\cite{Cheng87, Cheng90, Ecker88, Ecker92, Ecker94, BrunoPrades93}.
Direct photon emission can occur through both electric and magnetic dipole transitions.
The electric dipole transition can interfere with the IB amplitude giving rise to an interference term (INT), which can have CP violating contributions.
In ChPT, DE arises only at order $O(p^4)$ and cannot be evaluated in a model independent way.
The magnetic part is the sum of two anomalous amplitudes: one reducible, that can be calculated using the Wess-Zumino-Witten functional \cite{Wess:1971yu, Witten:1983tw}, and one direct amplitude, whose size is not model independent but is expected to be small.
The electric transition amplitude depends on unknown constants that cannot be determined by symmetry requirements alone.
An experimental measurement of both DE and INT terms allows the determination of both the electric and magnetic contributions.

The properties of the $\kppg$ decay can be conveniently described using the $\ecmk$, $W$ variables, where $\ecmk$ is the kinetic energy of the charged pion in the kaon rest frame and $W$ is a Lorentz invariant variable given by~\cite{christ,brib}:
\begin{eqnarray}
W^2=\frac{(P_{K} \cdot P_{\gamma})(P_{\pi} \cdot P_{\gamma})}{(m_Km_{\pi})^2}.
\end{eqnarray}
here $P_{K},P_{\pi},P_{\gamma}$ are the 4-momenta of the kaon, the charged pion and the radiative photon. Values of $W$ can vary within the range $0<W<1$.

Using these variables, the differential rate for the $\kppg$ process can be written as~\cite{christ, brib}:
\begin{eqnarray}
\frac{\partial^2 \Gamma^{\pm}}{\partial T_{\pi}^* \partial W}=\frac{\partial^2 \Gamma_{IB}^{\pm}}{\partial T_{\pi}^* \partial W}
\left[ 1+2\cos(\pm \phi+\delta_1^1-\delta_0^2) {m_\pi^2}{ m_K^2}
(X_E){ W^2} + \right. \nonumber \\
\left. {m_{\pi}^4}{m_{K}^4}(X_E^2+X_M^2){ W^4}\right],
\label{EQN_GPMW}
\end{eqnarray}
where $\frac{\partial^2 \Gamma_{IB}^{\pm}}{\partial T_{\pi}^* \partial W}$ is the differential rate for the IB component, $\phi$ is the CP violating phase, $\delta_l^I$ are the strong pion-pion re-scattering phases for a final state of isospin $I$ and orbital momentum $l$ of the $\pi \pi$ system, and $X_E$, $X_M$ are normalized electric and magnetic amplitudes respectively, which are W-independent quantities. Recently, the presence of a form factor in the pole part of the magnetic amplitude has also been suggested~\cite{dambrosio}.

The DE term is proportional to $W^{4}$ and the INT term is proportional to $W^{2}$.
This different $W$ dependence allows the extraction of the different decay components. In particular the ratio of the data $W$ distribution with respect to a simulation of the Inner Bremsstrahlung (MC(IB)) component can be parameterized as:
\begin{eqnarray}
Data/MC(IB)=c(1+(a\pm e)W^2+bW^4).
\label{EQN_dataMC}
\end{eqnarray}
The $c$ parameter represents an overall normalization factor, and the $a$, $b$ and $e$ coefficients are related to the branching ratios of Direct Emission and Interference respectively by:
\begin{eqnarray}
\frac{BR_{DE}}{BR_{IB}}=b \frac{\int dT_{\pi}^{*} \int  W^4 \frac{\partial^2 \Gamma_{IB}}{\partial T_{\pi}^{*}\partial W} \partial W}{\int dT_{\pi}^{*} \int \frac{\partial^2 \Gamma_{IB}}{\partial T_{\pi}^{*}\partial W} \partial W}= b \frac{I_{DE}}{I_{IB}},\\
\frac{(BR_{INT})_{CPC}}{BR_{IB}}=a \frac{\int dT_{\pi}^{*} \int  W^2 \frac{\partial^2 \Gamma_{IB}}{\partial T_{\pi}^{*}\partial W} \partial W}{\int  dT_{\pi}^{*} \int \frac{\partial^2 \Gamma_{IB}}{\partial T_{\pi}^{*}\partial W} \partial W}= a \frac{I_{INT}}{I_{IB}},\\
\frac{(BR_{INT})_{CPV}}{BR_{IB}}=e \frac{\int dT_{\pi}^{*} \int  W^2 \frac{\partial^2 \Gamma_{IB}}{\partial T_{\pi}^{*}\partial W} \partial W}{\int  dT_{\pi}^{*} \int \frac{\partial^2 \Gamma_{IB}}{\partial T_{\pi}^{*}\partial W} \partial W}= e \frac{I_{INT}}{I_{IB}}.
\label{EQN_brdeint_ab}
\end{eqnarray}

The coefficient $a$ parameterize the CP conserving part of INT term while the coefficient $e$ is the CP violating one. As it is expected that $a\gg e$, the CP violating contribution will be neglected in the study of the Interference term. In section \ref{par:W_CPV} the possibility of $e$ being non zero will be investigated.
Most of the previous experiments~\cite{Abrams_DE, Smith, Bolotov, Adler, Aliev_2000, Istra, Aliev_2005} have measured DE and INT terms in the kinematical region of 55~MeV $<\ecmk <$ 90~MeV, obtaining a value for the INT contribution compatible with zero. Therefore, the values quoted in these works for the DE branching ratio have been obtained from fits where the INT term has been set to zero. In the Particle Data Group (PDG) tables~\cite{PDG_08} only the latest experiments have been taken into account, resulting in $BR_{DE} = (4.3 \pm 0.7) \times 10^{-6}$.

\section{NA48/2 beam line and detector} \label{par:detector}
\begin{figure}[t]
\centering
    \includegraphics[width=14cm]{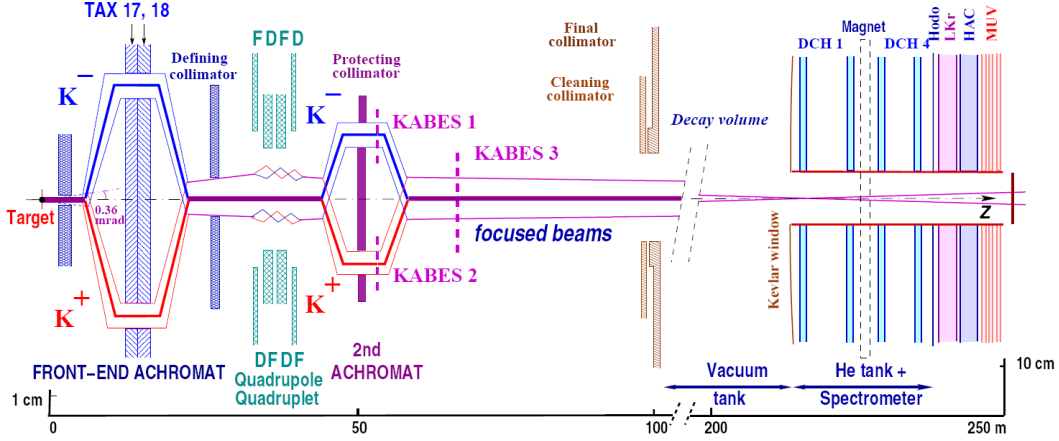}
\caption{Schematic lateral view of the NA48/2 beam line (TAX17,18: monitored beam dump/collimators used to select the momentum of the $K^+$ and $K^-$ beams; FDFD/DFDF: focusing set of quadrupoles, KABES1-3: beam spectrometer stations not used in this analysis), decay volume, and detector (DCH1-4: drift chambers, HOD: hodoscope, LKr: EM calorimeter, HAC: hadron calorimeter, MUV: muon veto). Note that the vertical scales are different in the two parts of the figure.}
\label{fig:beam}
\end{figure}

The beamline of the NA48/2 experiment was specifically designed to measure charge asymmetries in $\kppp$ decays~\cite{Batley:2006mu}, using secondary kaon beams produced by 400~GeV/{\it c} protons from the CERN SPS accelerator impinging on a Beryllium target.
The two simultaneous oppositely charged kaon beams with central momenta of ($60\pm3$) GeV/$c$ are selected by a system of dipole magnets forming a so-called ``achromat'' with null total deflection, followed by a set of focusing quadrupoles, muon sweepers and collimators (Figure~\ref{fig:beam}).
With $7\times10^{11}$ protons per burst of $\sim$ 4.5 s duration impinging on the target, the positive (negative) beam flux at the entrance of the decay volume is $3.8 \times 10^7$ ($2.6 \times 10^7$) particles per pulse, of which 5.7\% (4.9\%) are $K^+$ ($K^-$).
Downstream of the second achromat, both beams follow the same path, entering a decay volume housed in a 114 m long vacuum tank with a diameter of 1.92 m for the first 66 m, and 2.4 m for the rest. The beams are steered to be collinear within $\sim$ 1 mm in the entire decay volume.

A detailed description of the NA48 detector can be found in \cite{Fanti:2007vi}.
The charged decay products are measured by a magnetic spectrometer consisting of four drift chambers (DCH) with a dipole magnet placed in the middle.
Each octagonal shaped chamber has 4 views of sense wires, one horizontal, one vertical and two along each of two orthogonal 45$^{\circ}$  directions.
The spectrometer is located in a tank filled with helium at atmospheric pressure and separated from the decay volume by a thin (0.0031 radiation lengths, $X_0$) Kevlar window.
Downstream of the Kevlar window the beam continues in vacuum through a aluminium beam pipe of 152 mm outer diameter and 1.2 mm thick, traversing the center of the spectrometer and all following detectors.
Charged particles are magnetically deflected in the horizontal plane by an angle corresponding to a transverse momentum kick of 120 MeV/$c$.
The spectrometer momentum resolution is \textit{$\sigma_p /p = (1.02 \oplus 0.044\cdot p)\%$} ($p$ in GeV/$c$). The spectrometer is followed by a hodoscope consisting of two planes of plastic scintillators segmented into horizontal and vertical strips and arranged in four quadrants.
A quasi-homogeneous liquid Krypton calorimeter (LKr) is used to reconstruct $\gamma$ and electron showers.
It is an ionization chamber with an active volume of 7 m$^3$ of liquid krypton, segmented transversally into 13248 2 cm $\times$ 2 cm projective cells by a system of Cu-Be ribbon electrodes, and with no longitudinal segmentation.
The calorimeter is 27 $X_0$ deep and has an energy resolution $\sigma(E)/E = 0.032/\sqrt{E} \oplus 0.09/E \oplus 0.0042$ ($E$ in GeV).
The space resolution for single electromagnetic showers can be parametrized as $\sigma_x$ = $\sigma_y$ = $0.42/\sqrt{E} \oplus 0.06$ cm for each transverse coordinate $x$, $y$.
Due to the very good determination of the shower position the calorimeter allows a very precise reconstruction of the $\pi^0$ mass or of the Z coordinate of the decay vertex in  $\pi^0 \to \gamma\gamma$ decays.
To reduce the data volume only information from LKr cells with a signal greater than a given threshold is stored.
A hodoscope consisting of a plane of scintillating fibers is installed in the LKr calorimeter at a depth of $\sim$ 9.5 $X_0$.

NA48/2 collected data in two runs in 2003 and 2004.
In order to minimize systematic uncertainties for the asymmetry measurements, the magnetic fields in the spectrometer and beam line magnets were alternated regularly. The spectrometer magnet current was reversed on a daily basis during 2003 and every $\sim$ 3-4 hours in 2004. All the beam line magnet polarities were inverted once per week.

For one-track events, the first level trigger (L1) requires a signal in at least one quadrant of the scintillator hodoscope, in coincidence with the presence of energy depositions in the LKr geometrically consistent with more than two photons. At trigger level the signals from single cells are added together to form two orthogonal 4 cm wide views ($x$, $y$ projections).

The second level trigger (L2) is a software algorithm running on a fast on-line processor cluster. It receives the drift chamber information and reconstructs the momenta of charged particles.
Assuming that the particle is a $\pi^{\pm}$ originating from the decay of a 60 GeV/c $K^{\pm}$ traveling along the nominal beam axis, it evaluates the missing mass of the event. The requirement that the missing mass is not consistent with the $\pi^0$ mass rejects most of the main $\kpp$ background, reducing the rate of this trigger to $\sim$15k events per burst.

\section{Event selection}\label{par:selection}
A pre-selection is performed requiring events with one charged track of momentum above 10 GeV/$c$ and at least three electromagnetic clusters of energy greater than 3 GeV.
Cluster times must be within $3$~ns, and the track time within 4~ns of the mean cluster time.
This initial sample is kept for further analysis.

The track is considered to be a pion candidate if it lies within the spectrometer fiducial volume and the ratio between the energy deposited in the calorimeter and the momentum measured in the spectrometer ($E/p$) is smaller than 0.85.

Photon candidates are defined as in-time electromagnetic clusters reconstructed outside a 35 cm radius disk centered on the pion impact position at the front face of the LKr.
The number of $\gamma$ candidates must be three.
To avoid too large energy sharing corrections, the distance between any two photon clusters must be greater than 10 cm. Fiducial cuts on the distance of each photon from the LKr edges and center are also applied to ensure full containment of electromagnetic showers.

The closest distance of approach (CDA) between the charged pion trajectory and the beam axis is computed. A cut of CDA $<6$ cm is required, rejecting a negligible amount of the signal.
The position of the kaon decay vertex $z_{CH}$ is defined at the point of closest approach between these two lines.

Of the three selected photons, two of them must be associated to the $\pi^0$ decay and the remaining one identified as the radiated photon, also called the odd photon.
Using $z_{CH}$ the positions and energies of the photon clusters, the masses corresponding to the three possible pairings are calculated.
The pairing with mass closest to the PDG $\pi^0$ mass is then selected as the correct $\pi^0$ pairing.
For this pairing and using the PDG $\pi^0$ mass, the position of the decay vertex $z_{\pi^0}$ is calculated and must lie in a 9000~cm long decay volume starting 800~cm downstream from the final collimator.
Only events where $|z_{\pi^0}-z_{CH}|<400$~cm are selected, rejecting only $1\%$ of the signal.
In addition it is required that both remaining photon pairings in the event satisfy the condition $|z_{\pi^0}-z_{CH}|>400$~cm.
This reduces the data sample by $20\%$, but also reduces the misidentification of the odd photon to a level $<0.1\%$.

To reject background from channels with muons or from possible misreconstruction of charged tracks, the muon detector is required to have no hits.

Finally, the reconstructed kaon energy must be in the range $54<E_K<66$ GeV and the reconstructed kaon mass be within 10~MeV/$c^2$ of the PDG mass.

With this selection, L1 and L2 trigger efficiencies have been measured using control data samples recorded through minimum bias triggers.
The L1 trigger efficiency dependency on the cluster energies and relative positions has been studied using events with three photons in the calorimeter.
As a result, a cut has been added in the selection of $\kppg$ events, which requires a minimum distance between photon clusters in both the $x$ and $y$ coordinates, as seen by the trigger projections.
The L1 trigger inefficiency has been measured as a function of the minimum photon energies.
In order to exclude the region with smallest efficiency, a cut on the minimum photon energy of 5 GeV has been added to the event selection.
After rejecting periods affected by identified hardware problems,
the L1 trigger efficiency is measured to be greater than 99$\%$.

The effect of the L2 trigger on the $\ecmk$ distribution has been studied using data and Monte Carlo simulation.
The second level trigger applied an effective rejection of events with $\ecmk\gtrsim90$~MeV, designed to provide high efficiency for the $\kpppn$ mode and to suppress  $\kpp$ decays. In order to take into account on-line resolution effects and avoid relevant biases, the corresponding cut has to be tightened in the off-line selection of $\kppg$ to $\ecmk<80$~MeV. Data shows that the final L2 trigger efficiency is $> 97\%$ for all data taking periods, and is compatible with being flat in $W$.
MC simulation of the L2 trigger efficiency shows that it is flat in $W$ when a $\ecmk<80$~MeV cut is applied.
Any departure from flatness is observed at low $W$ only for events with $80<\ecmk<90$~MeV.

\section{Background contamination}\label{par:bacground}
\begin{figure}[h]
  \centering
  \includegraphics[width=8cm]{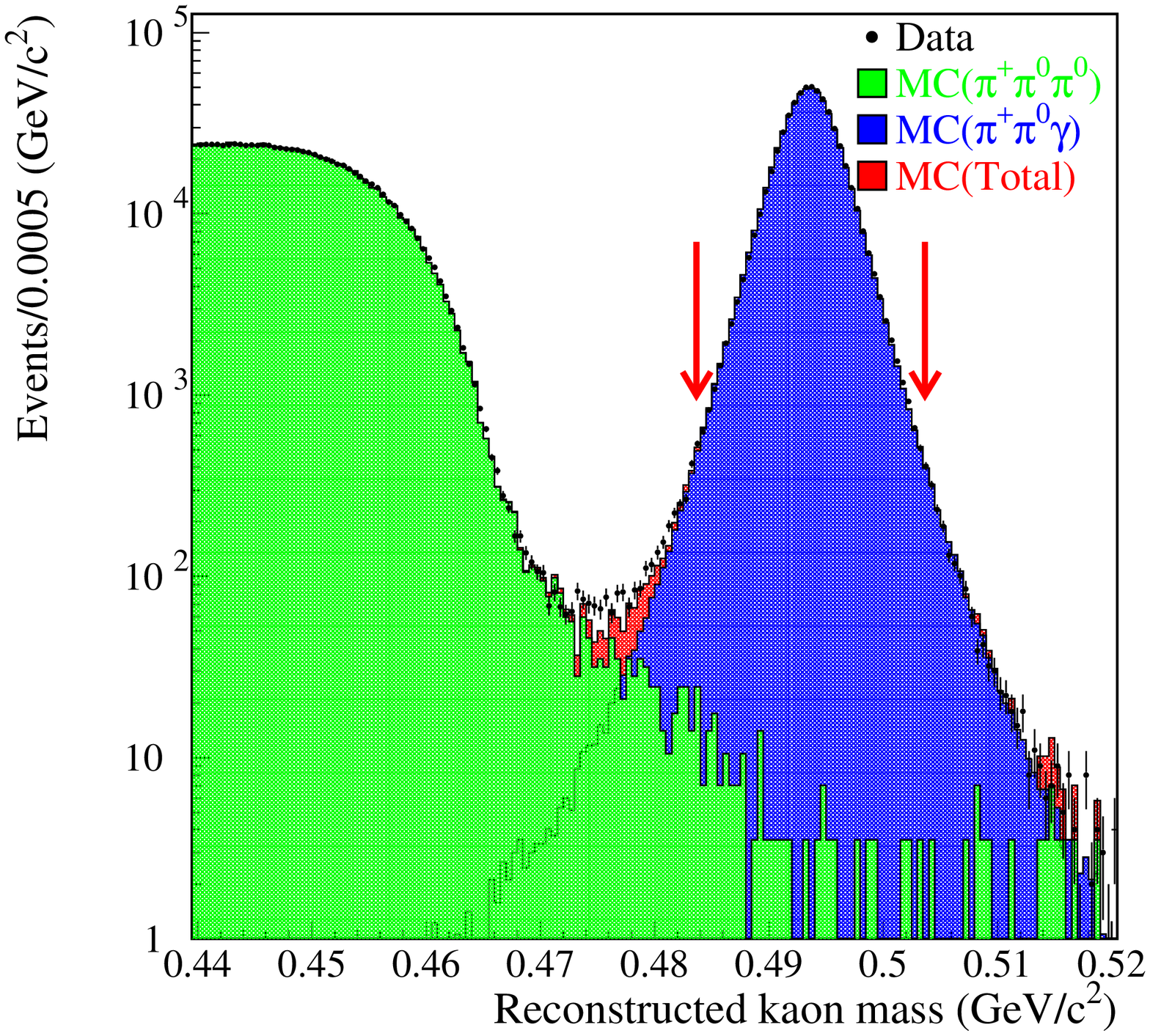}
  \caption{Mass distribution of $\kppg$ reconstructed candidates. Simulated $\kppg$ events are plotted in blue, while simulated $\kpppn$ events are superimposed in green. The data distribution (dots with error bars) is compatible with the sum of these two contributions, shown in red. Arrows delimit the selection region of $|m_{\pi^{\pm}\pio\gamma}- M_K| < 10~MeV/c^2$, where the amount of background from $\kpppn$ decays is negligible with respect to the total sample.}
  \label{fig:FIG_bg_final_Data_MC}
\end{figure}

Due to their large branching fractions and their particle content, the most important potential background contributions to $\kppg$ come from $\kpppn$ and $\kpp$ decays.
In most of the previous experiments these backgrounds have been eliminated by demanding $55<\ecmk<90$~MeV $\footnotemark[1]$.\footnotetext[1]{This cut is effective since $\ecmk(\kpp)\sim$110 MeV, while $\ecmk(\kpppn)<$53 MeV.}
As already mentioned, in NA48/2 a tighter cut of $\ecmk <$ 80~MeV has been implemented.
Any remaining background coming from $\kpp$ would manifest itself as an excess of events in the region $T^*_{\pi^0}\sim 110$~MeV, $T^*_{\pi^0}$ being the kinetic energy of the neutral pion in the kaon rest frame.
No significant excess of $\kpp$ events has been seen in the analyzed data sample, which is supported by the fact that data is compatible with signal in the high mass region (Figure~\ref{fig:FIG_bg_final_Data_MC}).

The lower $\ecmk>55$~MeV cut is very efficient against $\kpppn$ decays, but it also cuts away $\sim 50\%$ of the $\kppg$ DE component. In this analysis an alternative procedure has been used in order to suppress the $\kpppn$ background.
A $\kpppn$ event can be reconstructed as $\kppg$, either if one of the four final state photons from the two $\pi^0$s is undetected (being emitted outside of the acceptance or being very soft) or if two of the photons overlap in the LKr calorimeter.
These are mainly geometric effects and can be appropriately studied using Monte Carlo simulation.
$\kpppn$ decays with one very soft photon in the final state are strongly suppressed by phase space.

In order to reject background from $\kpppn$ decays with one undetected photon, two cuts have been implemented: the reconstructed kaon mass should be within $\pm$10~MeV/$c^2$ of the nominal kaon mass and the center of energy (COG) of the system should be within 2 cm of the beam line. The center of energy is defined as the energy weighted radial position, at the LKr calorimeter front plane, of the decay particles in the event, the charged track being projected onto the LKr using the track directions before the magnet deflection.

$\kpppn$ events with two photons overlapping in the LKr have the same signature as the signal in all detectors.
Moreover, the reconstructed values of kaon mass and COG satisfy the selection criteria, as there is no energy loss for this topology.

An algorithm has been developed to test the overlap hypothesis for all three photon clusters in the event.
Let us consider the three reconstructed photon clusters with energies $E_1$, $E_2$, $E_3$ and assume that the first one, with energy $E_1$, is really the overlap of two photons of energies $E =x E_1$ and $E^{\prime} = (1-x)E_1$.
Assuming that the photon with energy $E$ comes from the decay of the same $\pio$ as the second photon cluster of energy $E_2$, then the decay vertex for that $\pio$ would be:
\begin{equation}
z_{\pio}^1=\frac{\sqrt{(\mathrm{dist}_{1,2})^2 E E_2}}{M_{\pio}}=\frac{\sqrt{(\mathrm{dist}_{1,2})^2 xE_1 E_2}}{M_{\pio}},
\label{eqn:zpio1}
\end{equation}
where dist$_{1,2}$ is the radial distance between the photon clusters at the front of face of the LKr calorimeter, and $M_{\pio}$ is the nominal $\pio$ mass.\\
Similarly, assuming that the photon with energy $E^{\prime}$ comes from the decay of the same $\pio$ as the third cluster of energy $E_3$, the decay vertex for this second $\pio$ is given by:
\begin{equation}
z_{\pio}^2=\frac{\sqrt{(\mathrm{dist}_{1,3})^2 E^{\prime} E_3}}{M_{\pio}}=\frac{\sqrt{(\mathrm{dist}_{1,3})^2(1-x)E_1 E_3}}{M_{\pio}}.
\label{eqn:zpio2}
\end{equation}
As the two neutral pions originate from the same kaon decay, they must satisfy the constraint $z_{\pio}^1=z_{\pio}^2$$\equiv$ $z_{\pio}^{overlap}$.

Hence Equations~(\ref{eqn:zpio1}) and~(\ref{eqn:zpio2}) can be solved to obtain $z_{\pio}^{overlap}$.
The procedure is repeated for all three photon clusters.
The event is rejected if there is at least one solution in which $z_{\pio}^{overlap}$ is compatible with $z_{CH}$ within 400 cm.
This procedure rejects $\sim98\%$ of remaining $\kpppn$ background to be compared with $\sim99\%$ of $\ecmk >$55 MeV cut and allows placing the lower $\ecmk$ cut at zero, increasing the sensitivity to the DE and INT components.

Backgrounds from $K_{e3}$ and $K_{\mu3}$, also accompanied by either a radiative or an accidental photon, have been shown to be negligible using MC and the measured rate of accidental photons.

The mass distribution of the $\kppg$ candidates is shown in Figure~\ref{fig:FIG_bg_final_Data_MC}. All the background to $\kppg$ decays can be attributed to $\kpppn$ decays only, and is negligible in the selected mass region ($< 10^{-4}$ with respect to the signal).

\section{Simulation: resolution and acceptances}\label{par:mcsignal}

The NA48/2 Monte Carlo simulation is based on GEANT3~\cite{Geant}.
The $\kppg$ matrix elements for IB, DE and INT~\cite{brib,christ,dambrosio} have been implemented separately, so that the three terms could be studied independently.
No form factor for the DE contribution is considered at this stage.

The complete detector geometry and material description is included in the simulation.
The energy deposited in the detectors during tracking of the decay particles is digitized and the reconstruction is performed as for real data.
Detector imperfections, like DCH wire inefficiency maps, bad LKr calorimeter cells and malfunctioning readout cards are measured from data and included in the simulation in a run dependent way, so that experimental conditions are accurately reproduced.
This procedure allows the L2 trigger to be reliably simulated.

Non-gaussian tails on the cluster energy measurements may appear due to photon interaction with the LKr nuclei with a probability of $3 \times 10^{-3}$, resulting in an underestimate of the energy deposited in the electromagnetic shower.
These tails have been parameterized from data events and a correction has been implemented for simulated cluster energies, modeling their effect.

Using simulated events the relative resolution of $W$ has been found to be $\sim$ 1$\%$ in the range 0.2 $<W<$ 0.9.

To model the effect of the L1 trigger, the measured efficiency shape as function of the minimum photon energy in the event has been implemented in the simulation. 
After requiring the photon energy to be greater than 5 GeV (see section \ref{par:selection}) the effect of the L1 trigger was only significant for small $W$ values, at the level of 1--2$\%$.

After all cuts the final acceptances for IB, DE and INT were respectively 3.13 $\%$, 4.45 $\%$ and 4.21 $\%$ in the range 0.2 $<W<$ 0.9.

\section{DE and INT contributions} \label{sec:fitres}
The IB, DE and INT contributions have been extracted performing a fit of the data sample. Details on the fitting procedure, systematic uncertainties and final results are given in the following.
\subsection{Fitting procedure}
A fitting program has been developed, based on a Poissonian Maximum Likelihood (ML) method, for the extraction of the IB, DE and INT contributions present in the data sample.
The input information consists of the reconstructed $W$ distributions of data and simulated IB, DE and INT samples. For each of them, 14 bins have been considered in the range $0.2 < W < 0.9$.
The program calculates the relative contribution of each component by minimizing the difference between the number of data events and the resulting total number of simulated events in every $W$-bin. The fitted fractions are corrected for acceptance and final results are the relative contributions of DE and INT with respect to IB for $\ecmk$ $<$ 80~MeV and 0 $< W <$ 1:
\begin{eqnarray}
\mathrm{Frac_{DE}}= \frac{\mathrm{BR_{DE}}}{\mathrm{BR_{IB}}}=(3.32 \pm 0.15_{stat} \pm 0.14_{sys}) \times 10^{-2},
\label{EQN_resultDE} \\
\mathrm{Frac_{INT}}= \frac{\mathrm{BR_{INT}}}{\mathrm{BR_{IB}}}=(- 2.35 \pm 0.35_{stat} \pm 0.39_{sys}) \times 10^{-2}.
\label{EQN_resultINT}
\end{eqnarray}

The $W$ distributions used in the fit contain 599k data events, and 3.770M simulated events (3.339M IB, 220k DE, 211k INT) in the fitting range 0.2 $<$ $W$ $<$ 0.9.
The correlation coefficient between the DE and INT fractions is $-0.93$.

\begin{figure}[t]
$\begin{array}{c@{\hspace{0.2in}}c}
    \includegraphics[width=8.2cm]{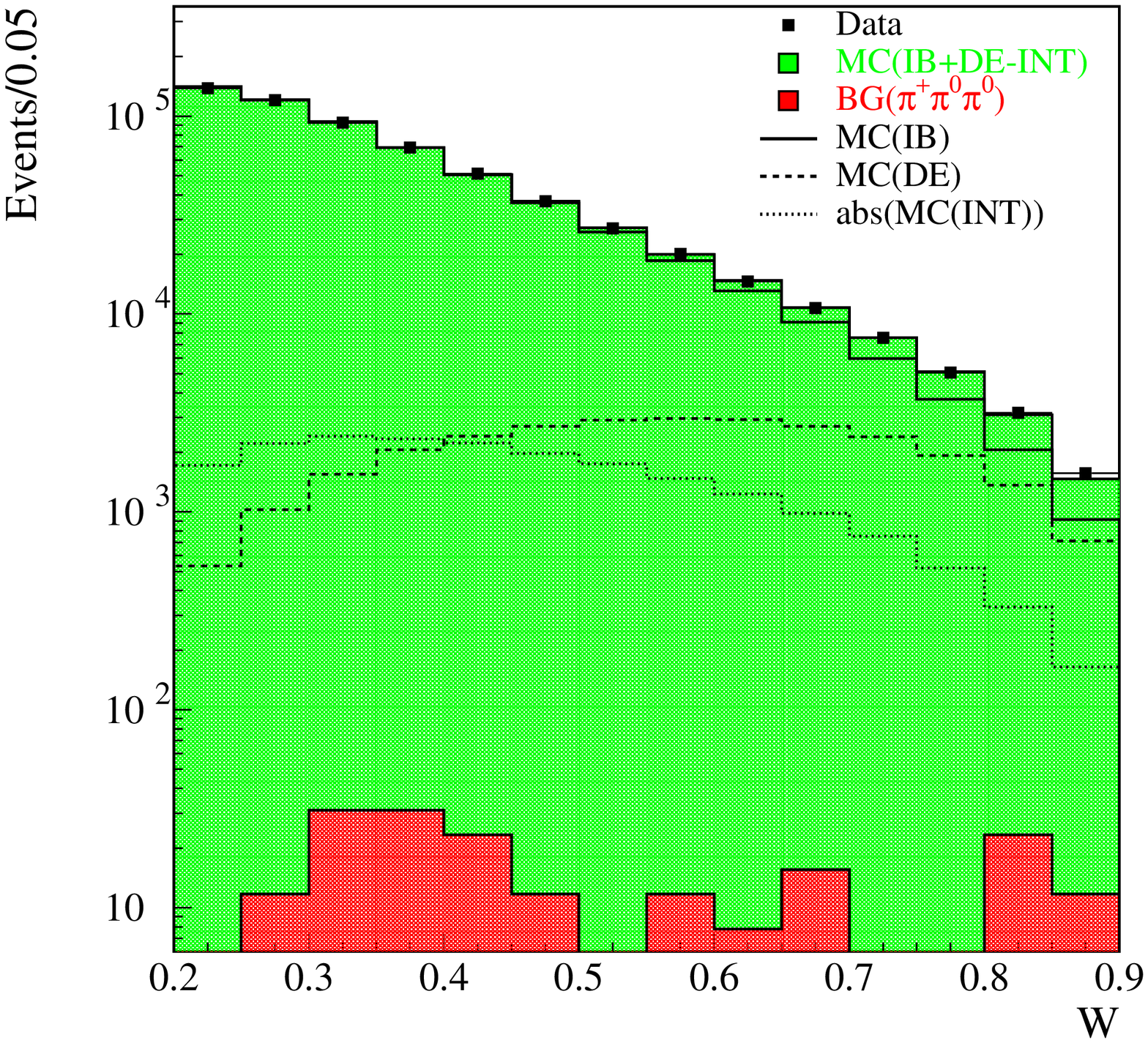} &
    \includegraphics[width=8 cm, height=8.2 cm]{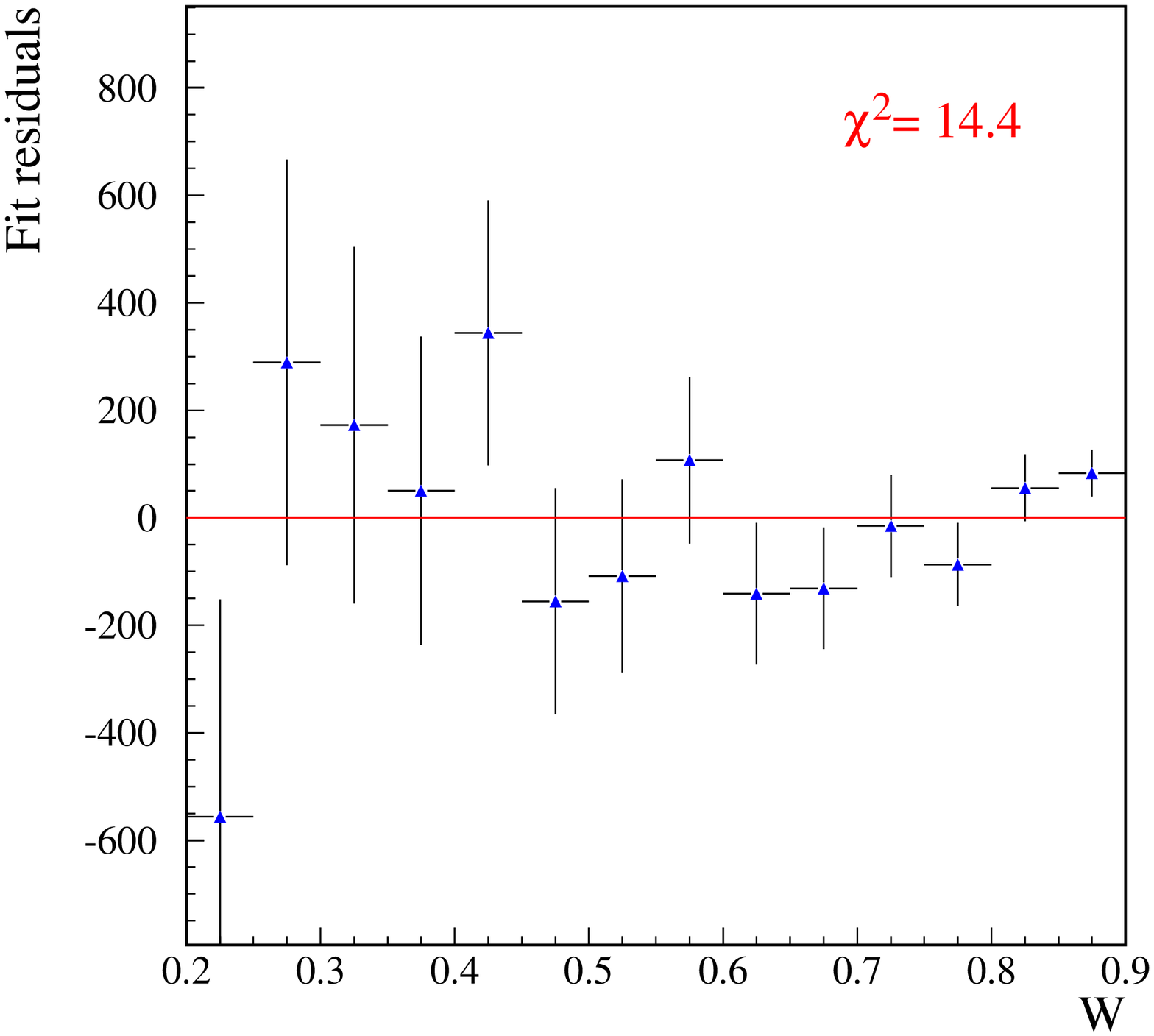}\\
\mbox{\bf (a)} & \mbox{\bf (b)}
\end{array}$
\caption{ (a) $W$ distribution for data (squares) and sum of simulated signal and background contributions (green).
The lines show the signal contributions: IB (continuous), DE (dashed), and absolute value of INT term (dotted). Background contribution is also shown (red).
(b) Residuals of the data with respect to the sum of simulated decay components weighted according to the Maximum Likelihood fit result.}
\label{fig:wsup}
\end{figure}

Figure~\ref{fig:wsup} a) shows the $W$ data distribution superimposed with the simulated IB, DE and INT $W$ distributions.
The background contribution is also shown. The residuals of the ML fit to the data are shown in Figure~\ref{fig:wsup} b). The $\chi^2$ of the residuals is 14.4 for 13 degrees of freedom corresponding to a probability of Prob($\chi^2$)=0.35.

\begin{figure}[t]
  \centering
    \includegraphics[width=8cm]{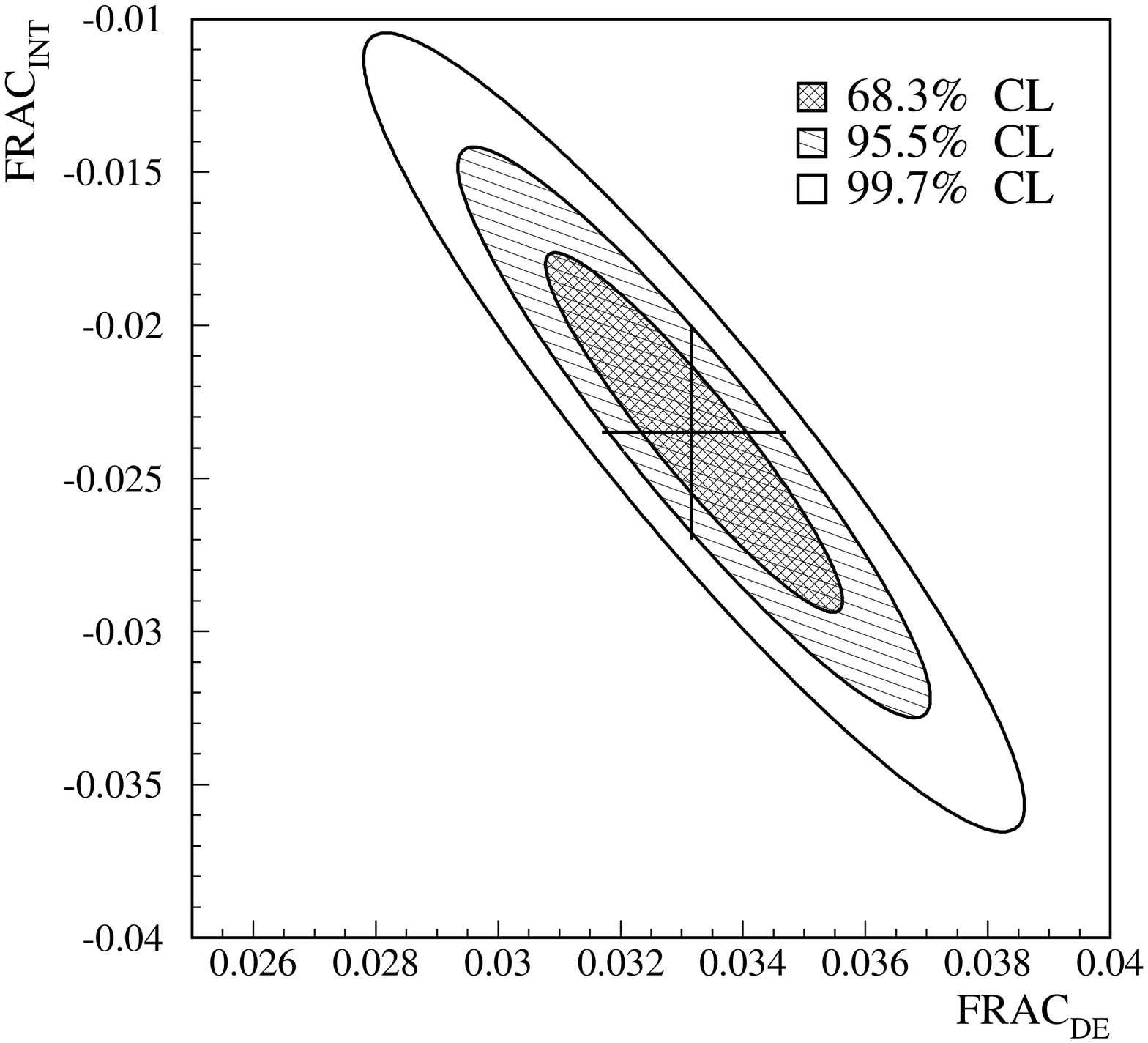}
  \caption{Contour plot for DE and INT terms. The cross shows the 1$\sigma$ statistical uncertainties on the projections.}
  \label{fig:FIG_cont_befbias}
\end{figure}

Contour plots have been computed requiring the logarithm of the likelihood to change by 1.15, 3.1 and 5.9 units. These correspond to probabilities in the DE-INT fractions plane of 68.3 $\%$, 95.5 $\%$ and 99.7 $\%$ respectively. They are shown in Figure~\ref{fig:FIG_cont_befbias}, including only statistical uncertainties.

\begin{figure}[t]
  \centering
    \includegraphics[width=9cm]{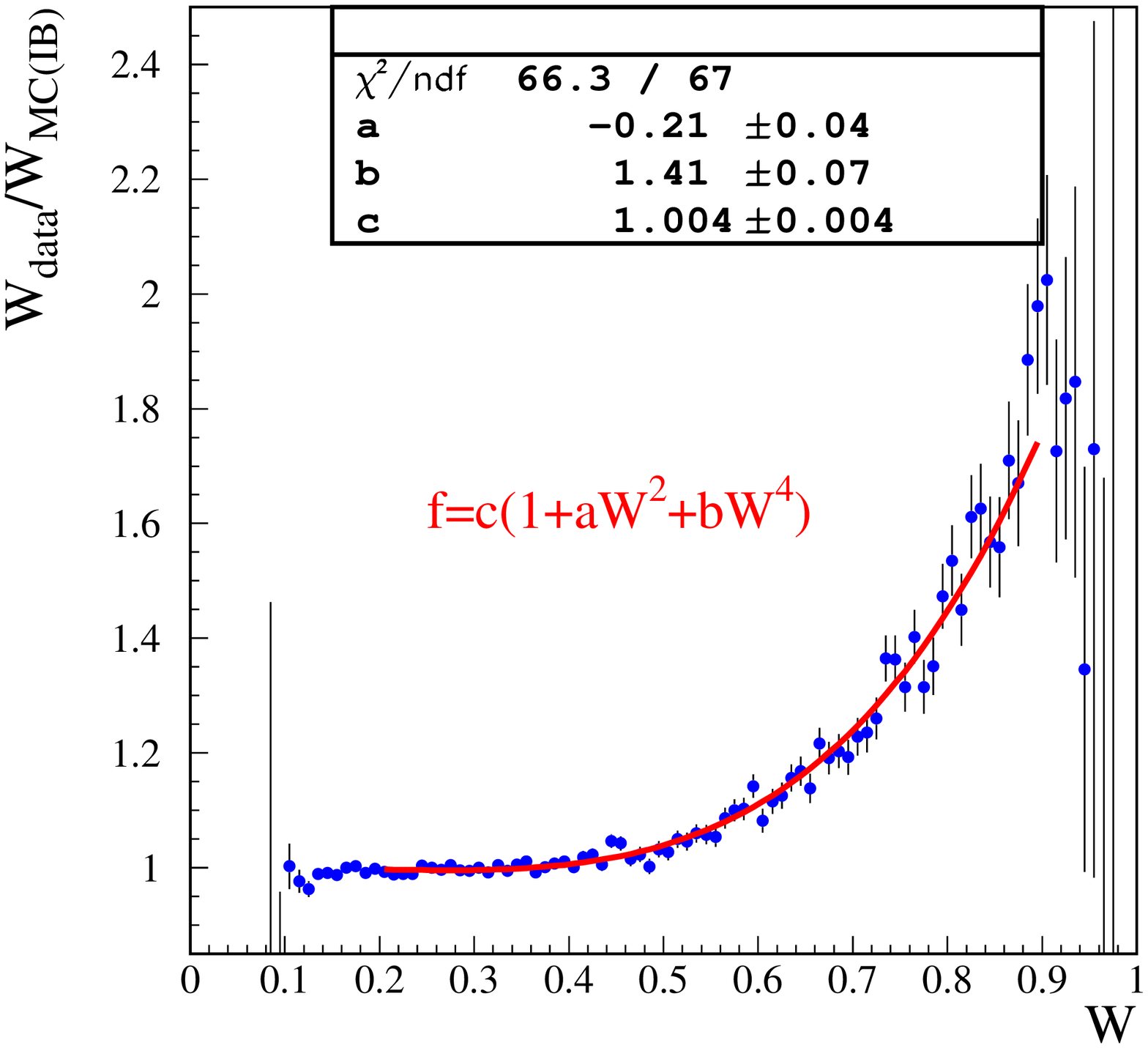}
  \caption{Ratio of $W$ distributions for data events with respect to IB simulated events. The red line shows the result of the fit to the polynomial form in Equation~\ref{EQN_dataMC} in the range  0.2$<W<$0.9.}
  \label{fig:fig_poly}
\end{figure}

Figure~\ref{fig:fig_poly} shows the ratio of the $W$ distributions of data and IB simulation after all selection cuts and corrections. For large values of $W$ the effect of the DE contribution is clearly seen.

A fit to this ratio can be used to determine $a$ and $b$ as defined in Equation~\ref{EQN_dataMC} and crosscheck the results obtained by the ML technique.
Neglecting differences in acceptance distributions for IB, DE and INT as a function of $W$, evaluating the necessary integrals over the Dalitz plot, and fitting in the range 0.2$<W<$0.9, the result is: $\mathrm{Frac_{DE}} =  (3.19 \pm 0.16_{stat}) \times 10^{-2}$ and $\mathrm{Frac_{INT}}= (- 2.21 \pm 0.41_{stat}) \times 10^{-2}$, which agrees with the result obtained with the ML method (Equations \ref{EQN_resultDE}, \ref{EQN_resultINT}) within the systematic uncertainty.

\subsection{Systematic uncertainties}
The stability of the maximum likelihood fit result with respect to acceptance knowledge, misreconstruction effects and residual background contamination has been checked by varying the values of the main selection cuts within a reasonable range.
The maximum likelihood fit has been repeated for every set of data and simulated samples obtained with the modified cut value, in order to study the differences in the resulting DE and INT fractions with respect to the standard (reference) ones.

\begin{itemize}

\item{ Acceptance control.

No significant effect was found varying the COG cut between 1 and 5.5 cm or the maximum distance between charged and neutral vertex
$|\Delta z| = |z_{\pio}-z_{\mathrm{CH}}|$ between 200 and 650 cm.
The minimum distance between the center of the photon cluster and the pion impact point at the LKr was varied between 20 and 50 cm, and the results are in agreement within the uncorrelated uncertainties.

The requirement that the photons cannot come from the vicinity of the DCH1 inner flange was released and no change was observed in the result within the uncorrelated uncertainties.
The acceptance definition of DCH 1, 2 and 4 was tightened by allowing the minimum and maximum radius of the track impact point to vary by a few cm away from standard cut values (12~cm and 150~cm).
The results for DE and INT fractions did not change within the uncorrelated uncertainties.
The distance between the neutral and charged vertex of the non-selected photon pairings was varied from 0 to 800 cm, giving statistically compatible results.

Limits on the contributions of the acceptance to the systematic uncertainties
were estimated to be $0.10\times10^{-2}$ for the DE term and $0.15\times10^{-2}$ for the INT term. These values
reflect the precision of the above studies involving the variation of the main
selection criteria defining the acceptance. A confirmation of this estimate is
given by the fact that the result of the a fit to the polynomial form (Equation~\ref{EQN_dataMC} and Figure~\ref{fig:fig_poly}), which uses
the acceptance in a slightly different way, does not deviate from the main ML result by more than the previously quoted systematic uncertainties.
}

\item{ Background control.

The $E/p$ requirement for the charged track was varied between 0.75 and 1, with negligible effect observed in the result.
A 10 GeV/$c$ cut is imposed on track momentum to maintain a high efficiency for the muon veto response, needed for background rejection. The value of this cut was varied between 5 and 15 GeV/$c$, giving statistically compatible results for DE and INT fractions.
The reconstructed kaon mass is required to be within $\pm$10~MeV/$c^2$ of the nominal kaon mass, corresponding to $\sim$ 4 standard deviations $\sigma$ of the resolution.
This value has been varied between 2.5 and 9.5 $\sigma$. It is expected that releasing this cut allows background from $\kpppn$ decays to enter the sample. However, the background is well enough separated from the signal, and no effect can be seen up to $\sim$ 7 $\sigma$. Therefore no systematic uncertainty was assigned due to background contamination.
}
\end{itemize}
In addition to varying the relevant cuts, other effects have been studied.
\begin{itemize}

\item{Trigger efficiency.

The effect of the L1 trigger correction has been evaluated performing 1000 different fits. In each fit the bin values of the correction have been randomly varied according their uncertainties. From the distributions of fit results, rms values of 0.01$\times10^{-2}$ for DE and 0.03$\times10^{-2}$ for INT have been extracted.

The effect of a possible residual L2 trigger inefficiency has been studied changing the upper $\ecmk$ limit from 80~MeV to 65~MeV  in five steps. To make a comparison with the standard measurement, all results have been extrapolated to the $\ecmk$ $<$ 80~MeV region. A significant change in the result is observed only for the INT term. This behavior is in agreement with the hypothesis of a resolution effect of the L2 trigger response. The maximum observed difference, 0.3, has been assigned as systematic uncertainty to the INT result.}
\item{LKr energy reconstruction.

The absolute LKr energy scale is known to a precision of 0.1 $\%$. In order to evaluate the effect of this uncertainty, the data cluster energies have been multiplied by 1.001. The results of the ML fit changed by 0.09$\times10^{-2}$ for the DE fraction and by 0.21$\times10^{-2}$ for the fraction of INT. This difference has been assigned as systematic uncertainty to the result.

Due to the zero-supression threshold applied to the LKr calorimeter cells at the readout, a non linear relation develops between the value of the energy deposited in the calorimeter and its actual measurement. This non linear response is relatively more important for small cluster energies. Using $\kspiopio$ decays from the 2002 run a correction has been applied to clusters with energies smaller than 11 GeV.
This can be used for 2003 and 2004 data, where the readout thresholds were kept at the same value as in 2002.
The shape of this non-linearity correction has been changed within a reasonable range.
The results of the ML fit did not change with respect to the reference within the uncorrelated uncertainties.}

\item{Fitting method.

Combinations of independent simulated samples of IB, DE and INT have been used as fake data, and fitted against standard simulated samples used in the data fit. In this way possible biases in the fitting technique have been excluded.}

\item{Radiative corrections.

The simulation of the Inner Bremsstrahlung component has been interfaced with the PHOTOS package~\cite{photos} thus generating one or more radiative photons.
Only 1$\%$ of the selected events in this sample had more than one radiated photon. The ratio of the $W$ distribution of these multi-photon events to the $W$ distribution of the standard IB events was found to be a linear function of $W$.
This model was used in a simple Monte Carlo program to evaluate its impact on the DE and INT fractions. This was found to be much smaller than $0.01\times10^{-2}$ and therefore negligible.
}
\item{Resolution control.

Differences in resolution between data and simulation can produce different distortions of the $W$ shape, potentially biasing the DE and INT measurements.
The comparison of kaon and neutral pion mass distributions in data and simulation shows relative resolution differences smaller than 2$\%$.
In addition, $W$ distributions of simulated events have been smeared by different amounts and used as fake data to be fitted against standard Monte Carlo samples with no extra smearing.
These studies show that the systematic bias on the extraction of the DE and INT fractions is negligible if the difference of $W$ relative resolution between data and MC is smaller than 5$\%$.}
\end{itemize}
The main contributions to the systematic uncertainty are summarized in Table~\ref{systematics}.
\begin{table}[h!]
\begin{center}
\begin{tabular}{| l | r| r |}
\hline
Source of systematic uncertainty     & Effect on DE term    & Effect on INT term\\
\hline
 Detector acceptance      &  $<0.10\times10^{-2}$    & $<0.15\times10^{-2}$ \\
 L1 trigger efficiency    &  $0.01\times10^{-2}$     & $0.03\times10^{-2}$  \\
 L2 trigger efficiency    &  $0.00\times10^{-2}$     & $0.30\times10^{-2}$  \\
 LKr Energy Scale         &  $0.09\times10^{-2}$     & $0.21\times10^{-2}$  \\
 \hline
 Total systematic uncertainty     &        $0.14\times10^{-2}$    & $0.39\times10^{-2}$          \\
 \hline
\end{tabular}
\caption{Summary of non negligible systematic uncertainties.\label{systematics}}
\end{center}
\end{table}

The effects of a possible form factor on the DE and INT terms measurements
have been investigated. Following the work of reference~\cite{dambrosio}, a possible form
factor in the pole part of the magnetic amplitude can be parameterized by the
quantity $\eta_V$. The presence of such a form factor, if neglected in the
analysis, induces anti-correlated variations of the extracted DE and INT terms such that
$\Delta \mathrm{Frac_{DE}} = -\Delta \mathrm{Frac_{INT}}= -0.01\eta_V$
Therefore the correction to the values of the DE and INT terms can be easily
evaluated for any value of $\eta_V$ in the allowed range [0,1.5]. One should
note that neglecting in the analysis the existence of a non-zero form factor present in the data, only a positive INT term could be induced, in no way faking a negative INT term.

\subsection{Results and discussion}
As seen from Equation~\ref{EQN_GPMW}, the simultaneous measurement of the DE and INT terms in $\kppg$ decays allows to quantify the electric $X_E$ and magnetic $X_M$ contributions.
Assuming a negligible amount of CP violation in the $\kppg$ decays, the $\phi$ angle can be set to zero.
The cosine of the difference between the two strong re-scattering phases can be approximated to 1 (see section \ref{par:CPV}).
The values obtained for $X_E$ and $X_M$ are
\be
X_E = (- 24 \pm 4_{stat} \pm 4_{sys}) \: \rm{GeV}^{-4}, 
\label{EQN_xe}
\ee
\be
X_M = (254 \pm 6_{stat} \pm 6_{sys}) \: \rm{GeV}^{-4}.
\label{EQN_xm}
\ee
The correlation coefficient is $-0.83$.

The hypothesis that the chiral anomaly is the only source of magnetic amplitudes in the Direct Emission term predicts $X_M \sim 270~$GeV$^{-4}$~\cite{Ecker92, D'Ambrosio:1994du}.
Factorization models predict both an enhancement of $X_M$ with respect to the pure chiral anomaly calculation and a positive $X_E$ term in $\kppg$~\cite{Ecker92, Ecker94, D'Ambrosio:1994ae}.
Our result shows a magnetic part compatible with the pure chiral anomaly and a negative $X_E$ term, indicating that factorization models cannot provide an appropriate description of DE and INT in $\kppg$.

In order to compare the NA48/2 results with those from previous measurements, the ML fit of the selected sample has been redone setting the interference term to zero, and the result for DE extrapolated to 55 $<\ecmk<$ 90~MeV.

\begin{figure}[t]
  \centering
    \includegraphics[width=8.2cm]{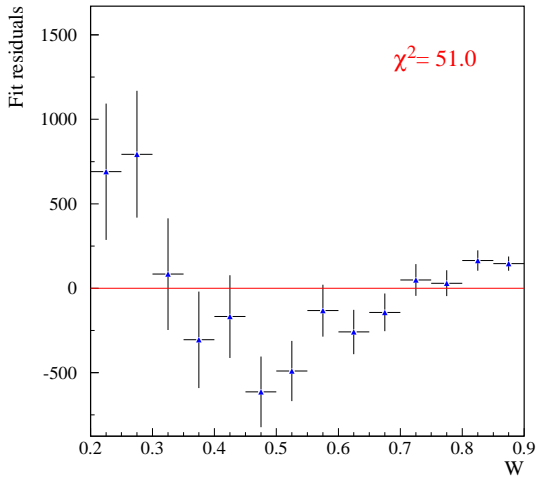}
  \caption{Residuals of the Maximum Likelihood fit to the data with the INT contribution set to zero, with respect to the simulated distribution obtained from the weighted sum of the IB and DE components.}
  \label{fig:res2par}
\end{figure}

Figure~\ref{fig:res2par} shows the fit residuals.
The $\chi^2$ demonstrates that the data distribution cannot be properly described without an interference term and that the DE-only fit is not appropriate for this data.
The result of this fit extrapolated to 55 $<\ecmk<$ 90~MeV is given here just for completeness:
\begin{equation}
\mathrm{Frac_{DE} (INT=0)}= (0.89 \pm 0.02_{stat} \pm 0.03_{sys}) \times 10^{-2}.
\end{equation}
The systematic uncertainty assigned to this measurement corresponds to half of the difference observed in the result varying the $\ecmk$ upper cut from 80~MeV to 65~MeV.

Using the theoretical prediction for the branching ratio of the inner bremsstrahlung component as given in~\cite{brib} and the current value for the branching ratio of the direct emission component as quoted in the PDG \cite{PDG_08} one obtains $\mathrm{Frac_{DE} (INT=0)}= (4.3 \pm 0.7) \times 10^{-6}/(2.61 \times 10^{-4})= (1.65 \pm 0.27) \times 10^{-2} $.

It should be noted that NA48/2 has been able to keep the rate of wrong solutions for the odd photon to $<0.1\%$ while in other measurements ~\cite{Abrams_DE, Smith, Bolotov, Adler, Aliev_2000, Istra, Aliev_2005} the wrong solutions fraction was always larger than $10\%$. The relative background contamination is $<10^{-4}$ while in other experiments it is at the level of $10^{-2}$ or larger. The statistics used for this measurement is more than one order of magnitude larger than the sum of all previous experiments. In addition this measurement uses for the first time data without any lower cut on $\ecmk$, with the benefit of better sensitivity to both DE and INT contributions to the $\kppg$ decay amplitude.

\section{CP violation}\label{par:CPV}

As the decay of $\kppg$ with direct photon emission is not suppressed by $\Delta I=1/2$ rule, it has always been considered a good channel to search for CP violation \cite{brib,kubir}.
According to \cite{Cheng:1993xd} the magnitude of CP asymmetry in the Dalitz plot ranges from $2\times10^{-6}$ to $1\times10^{-5}$ when the center of mass photon energy varies from 50 MeV to 170 MeV.
Possible supersymmetric contributions to direct CP violation in kaon decays can push the asymmetry to the level of $10^{-4}$ in a specific region of the Dalitz plot \cite{cpv_isidori}.
Present experimental knowledge on the asymmetry dates back to the late seventies and is $(0.9 \pm 3.3)\%$ \cite{PDG_08}.


In order to measure CP asymmetry in $\kppg$ decays, the previously described selection (Section~\ref{par:selection}) has been modified to increase
the statistical precision by setting the minimum photon energy to 3 GeV and eliminating the $W>0.2$ cut.

The new selection preserves the performance in terms of background rejection. Even if the selection criteria have not been completely optimized for studies of CP violating effects, it must be noted that the design of the experiment is such as to suppress beam and detector related differences between $K^+$ and $K^-$ decays.

For this reason differences for the two kaon charges in trigger efficiencies and acceptances can be neglected, as proven in \cite{Batley:2006mu}.
The residual effects are taken into account in the systematic uncertainty.
To investigate CP violation in $\kppg$, events were analyzed according to the reconstructed kaon charge, leading to a sample with 695k $K^+$ and 386k $K^-$.

\subsection{Integrated charge asymmetry}

The simplest observable that can be measured is the difference in the decay rates of $K^+$ and $K^-$. This can be expressed as the asymmetry on the total number of events A$_N$ defined as:
\begin{equation}
A_N=\frac{N_+ - R N_-}{N_+ + R N_-}
\label{eqn:AN}
\end{equation}
where $N_+$, $N_-$ are the number of $K^+$, $K^-$ decays to $\pi^{\pm} \pio \gamma$ in the data sample, and $R$ is the ratio of the number of $K^+$ to $K^-$ in the beam.
Since $R$ cannot be directly measured in NA48/2, normalization to another decay channel is required.
For this work, the $\kpppn$ decay has been chosen as normalization.
The previously measured CP asymmetry in the rate for this channel is compatible with zero ($\frac{\Delta\Gamma}{\Gamma}=(0.0\pm0.6)$ $10^{-2}$~\cite{PDG_08}), and its uncertainty would contribute as an external source of uncertainty to our measurement of the CP rate asymmetry in the $\kppg$ channel.
However, the NA48/2 data has shown no CP violation asymmetry in the Dalitz plot for $\kpppn$ decays at the level of 10$^{-4}$ \cite{Batley:2006mu}. Since in most models the integrated rate asymmetries are expected to be smaller than the slope asymmetries (see for example~\cite{IsiMaiPug}), we consider the absence of a CP rate asymmetry in the $\kpppn$ channel at the same level a plausible assumption, making the above external error negligible.

A large number of $\kpppn$ decays has been collected by NA48/2. Using the selection described in \cite{Batley:2005ax}, only adapting
few geometrical cut to fit the $\kppg$ ones, a measurement of the ratio $R=1.7998\pm0.0004$ has been performed with a high accuracy of $\delta R/R\sim 2\times10^{-4}$.

The ratio $R$ has been computed in bins of kaon momentum and the corresponding asymmetry calculated.
No dependency of the asymmetry was found neither as a function of $\ecmk$ nor as a function of the kaon momentum spectra.
Differences in $\pi^+$-proton and $\pi^-$-proton cross section can induce a difference in the L1 trigger efficiencies for $K^+$ and $K^-$ events, that can reflect in a fake asymmetry in $\kppg$.
The charged hodoscope inefficiency in detecting a pion depends on the pion hadronic cross section, and is different for the two charges.
However, first order effects are also included in the $R$ measurement performed using $\kpppn$.
The major difference between $\kppg$ and $\kpppn$ is due to different pion momentum spectra.
The effect on the asymmetry has been calculated to be of the order of a few 10$^{-5}$.
A possible reconstruction induced asymmetry has been obtained to be $<5\times10^{-4}$ using MC, while trigger effects have been evaluated to be of the order of $4\times10^{-4}$.
Finally, including the maximum $R$ variation allowed by the NA48/2 estimation, the measured asymmetry is:
\begin{eqnarray} A_N=(0.0\pm 1.0_{stat}\pm 0.6_{sys})\times 10^{-3}. \end{eqnarray}

From the above value a limit for the rate asymmetry of $|A_N| < 1.5\times 10^{-3}$ at 90$\%$ CL can be deduced.
An alternative approach would have been to normalize the $K^+$ and $K^-$ fluxes using inclusive decay channels, such as the inclusive three pion channels or the $\kpp$ plus $\kppg$ channels. For those the absence of a CP asymmetry is guaranteed by the CPT theorem together with unitarity when small final state interactions are neglected~\cite{DCline,YUeda}.
However, in the NA48/2 experiment other possible normalization channels different from $\kpppn$ have been collected with a trigger different from that used for $\kppg$ decays, so that using them in the asymmetry measurement would require an accurate knowledge of the relative trigger efficiencies. Therefore such an alternative is not presented at this time.

The asymmetry $A_N$ can be related to the CP violating phase $\phi$ by Equation~\ref{EQN_GPMW}:
\begin{eqnarray}
A_{N}&=&\frac{\Gamma^+ - \Gamma^-}{\Gamma^+ + \Gamma^-} \sim \frac{\Gamma^+ - \Gamma^-}{2 \Gamma_{IB}}=(I_{INT}/I_{IB})2X_Em_K^2 m_{\pi}^2 \sin \phi \sin(\delta_1^1-\delta_0^2)= \nonumber \\
     &=&e(I_{INT}/I_{IB}), \\
\sin \phi&=&\frac{A_{N}}{(I_{INT}/I_{IB})2X_Em_K^2 m_{\pi}^2\sin(\delta_1^1-\delta_0^2)},
\end{eqnarray}
where the parameter $e$, describing the asymmetry, has been introduced.

The difference between the two strong re-scattering phases $\delta_1^1$, $\delta_0^2$ is evaluated to be $6.6^{\circ} \pm 0.5^{\circ}$.
This is calculated using the theoretical predictions for $\delta_1^1$ and $\delta_0^2$ as a function of the $\pi\pi$ mass from \cite{brigitte}, weighted by the observed $M_{\pi+\pi0}$ data distribution.

Using in addition the value of 0.105 for the ratio of $W$ integrals, and $X_E$ from the NA48/2 measurement (Equation \ref{EQN_xe}) a measurement for $\sin \phi=-0.01 \pm 0.43$,  or $|\sin \phi|< 0.56$ at 90$\%$ CL is obtained.

\subsection{Fit to the $W$ spectrum} \label{par:W_CPV}
Another interesting check of CP violation can be obtained by looking at the distribution of the asymmetry as function of the Dalitz plot variable $W$.
In fact, an enhancement of the asymmetry in particular regions of the Dalitz plot is suggested by \cite{Cheng:1993xd, cpv_isidori}. Using Equation~\ref{EQN_GPMW}, the following $W$ dependence of the asymmetry ($A_W$) is predicted:
\begin{eqnarray}
\frac{d\Gamma^{\pm}}{dW}&=&\frac{d\Gamma_{IB}}{dW}(1+(a\pm e)W^2+bW^4), \\
\frac{dA_W}{dW}&=&\frac{\frac{d\Gamma^{+}}{dW}-\frac{d\Gamma^{-}}{dW}}{\frac{d\Gamma^{+}}{dW}+\frac{d\Gamma^{-}}{dW}}=\frac{eW^2}{1+aW^2+bW^4},
\label{eqn:ANW}
\end{eqnarray}
where $a$ and $b$ have been already extracted in section \ref{sec:fitres}, neglecting CP violation effects.
\begin{figure}[t]
  \centering
  \includegraphics[width=8.2 cm]{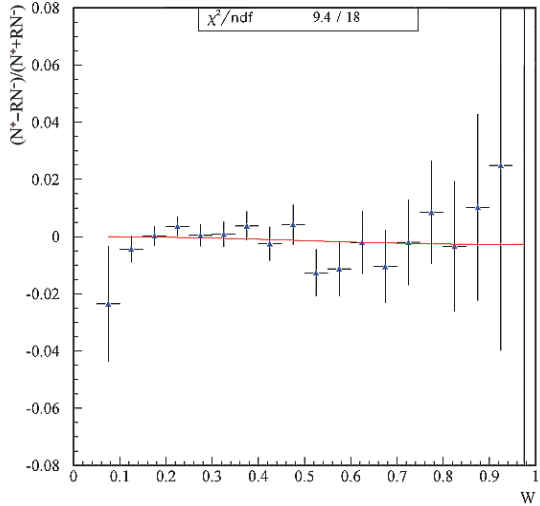}
  \caption{Measured asymmetry as defined in Eq. \ref{eqn:ANW} as a function of $W$.} \label{fig:Wasym}
\end{figure}
The parameter $e$ is the only free parameter left in the fit. Figure~\ref{fig:Wasym} shows the asymmetry as a function of the $W$ variable for the whole data sample together with the fit result.
Multiplying the $e$ parameter obtained from the fit with the value of the integral ratio $I_{INT}/I_{IB}$ = 0.105, the value of $A_W$, the asymmetry in the $W$ spectra, is:
\begin{eqnarray}
A_W = e I_{INT}/I_{IB}=(- 0.6 \pm 1.0_{stat})\times 10^{-3}.
\end{eqnarray}
The value of $A_W$ is compatible with the result $A_N$ of the overall charge asymmetry.

\section{Summary}
From a sample of about 600k $\kppg$ decay candidates, the NA48/2 experiment has measured the relative amounts of DE and INT with respect to the internal bremsstrahlung (IB) contribution in this decay in the range $0<\ecmk<80$~MeV:
\begin{center}
$\mathrm{Frac_{DE}}{(0<\ecmk<80~\mathrm{MeV})}=(3.32\pm0.15_{stat}\pm0.14_{sys}) \times 10^{-2}, $ \\
$\mathrm{Frac_{INT}}{(0<\ecmk<80~\mathrm{MeV})}= (- 2.35\pm0.35_{stat}\pm0.39_{sys}) \times 10^{-2}. $ \\
\end{center}
The relative background contamination has been kept to $<10^{-4}$, and the rate of wrong solutions for the odd photon to $<0.1\%$. Thanks to the implementation of an algorithm rejecting background from $\kpppn$ decays, the cut on $\ecmk$ could be released below the standard $55$~MeV used by most of the previous experiments, gaining in sensitivity to both DE and INT contributions to the $\kppg$ decay amplitude.
This measurement constitutes the first observation of an interference term in $\kppg$ decays.
The results for electric and magnetic contributions, $X_E = (- 24 \pm 4_{stat} \pm 4_{sys}) \: \rm{GeV}^{-4}$ and $X_M = (254 \pm 6_{stat} \pm 6_{sys}) \: \rm{GeV}^{-4}$ indicate that factorization models cannot be applied to estimate the size of DE and INT in $\kppg$ and that the magnetic part is compatible with pure chiral anomaly.

Using a slightly modified event selection, two samples of 695k $K^+$ and 386k $K^-$ have been reconstructed and used to set a limit on the CP violating asymmetry in the $K^+$ and $K^-$ branching ratios for this channel of less than 1.5 $\times 10^{-3}$ at 90$\%$ confidence level.
For this measurement the $\kpppn$ decay has been used for normalization.

\section*{Acknowledgements}
It is a pleasure to thank the technical staff of the participating laboratories, universities and affiliated computing centers for their efforts in the construction of the NA48 apparatus, in the operation of the experiment, and in the processing of the data. We would also like to thank G. D'Ambrosio and G. Isidori for constructive discussions.

\newpage

\end{document}